\documentclass[usegraphicx,usenatbib]{mn2e}

\begin{document}

\voffset -0.5in


\title[Interstellar C$_2$ and C$_3$ in SMC]{Detection of Interstellar C$_2$ and C$_3$ in the Small Magellanic Cloud\thanks{Based on observations collected at the European Southern Observatory, Chile, under program 382.B-0556.}}

\author[Welty et al.]{Daniel E. Welty$^1$, J. Christopher Howk$^2$, Nicolas Lehner$^2$, and John H. Black$^3$ \\
$^1$University of Chicago, Astronomy \& Astrophysics Center, 5640 S. Ellis Ave., Chicago, IL, USA; dwelty@oddjob.uchicago.edu \\
$^2$University of Notre Dame, Department of Physics, Center for Astrophysics, Notre Dame, IN, USA; jhowk@nd.edu, nlehner@nd.edu \\
$^3$Onsala Space Observatory, Chalmers Inst. of Technology, Onsala, SE-439 92, Sweden; John.Black@chalmers.se}

\maketitle
\begin{abstract}

We report the detection of absorption from interstellar C$_2$ and C$_3$ toward the moderately reddened star Sk~143, located in the near 'wing' region of the SMC, in optical spectra obtained with the ESO VLT/UVES.
These detections of C$_2$ (rotational levels $J$=0--8) and C$_3$ ($J$=0--12) absorption in the SMC are the first beyond our Galaxy.
The total abundances of C$_2$ and C$_3$ (relative to H$_2$) are similar to those found in diffuse Galactic molecular clouds -- as previously found for CH and CN -- despite the significantly lower average metallicity of the SMC.
Analysis of the rotational excitation of C$_2$ yields an estimated kinetic temperature $T_{\rm k}$ $\sim$ 25 K and a moderately high total hydrogen density $n_{\rm H}$ $\sim$ 870 cm$^{-3}$ -- compared to the $T_{01}$ $\sim$ 45 K and $n_{\rm H}$ $\sim$ 85--300 cm$^{-3}$ obtained from H$_2$.
The populations of the lower rotational levels of C$_3$ are consistent with an excitation temperature of about 34 K.

\end{abstract}



\section{INTRODUCTION}
\label{sec-intro}

Understanding the properties of diffuse molecular gas, where hydrogen makes the transition from being predominantly atomic to predominantly molecular, is key for understanding the formation of molecular clouds -- with consequent implications for both star formation and galactic evolution.
Moderately reddened sight lines in the Magellanic Clouds can probe such diffuse molecular gas under environmental conditions somewhat different from those found in the local Galacic interstellar medium (ISM).
Both the metallicities and the average dust-to-gas ratios are lower in the Magellanic Clouds -- by factors of 2--3 in the Large Magellanic Cloud (LMC) and 4--5 in the Small Magellanic Cloud (SMC) (e.g., Smith 1999; see also references in appendices to Welty et al. 1997, 1999; Welty, Xue, \& Wong 2012).
The ambient UV radiation fields are (on average) stronger by factors of a few (e.g., Lequeux 1989), and the fields within the interstellar clouds are further modified by differences in UV extinction, especially in the SMC (e.g., Gordon et al. 2003; Cartledge et al. 2005).
Such environmental differences are predicted to significantly affect both the structure and the composition of diffuse molecular clouds.
For example, the photon-dominated regions surrounding largely molecular cores are expected to be more extensive in the Magellanic Clouds, as are regions of so-called 'dark' molecular gas, with significant amounts of H$_2$ not traced by CO emission (Maloney \& Black 1988; Pak et al. 1998).
The interplay between the lower metallicities, differences in relative elemental abundances, and such structural differences will likely yield both obvious and subtler differences in the abundances of various molecular species, in both relatively diffuse and denser clouds (e.g., Johansson 1997).
Studies of these two nearby low-metallicity systems -- where we know the stellar abundances and can compare emission and absorption diagnostics in many locations -- thus can provide insights into both interstellar chemical processes and the conditions in more distant low-metallicity galaxies, where the systems are poorly resolved and the information is not as complete.

Large-scale surveys of the CO emission in the Magellanic Clouds, undertaken with the 4-m NANTEN telescope at a resolution of 2.6 arcmin, have indicated the overall distributions of the denser molecular gas in the LMC and SMC (Mizuno et al. 2001; Yamaguchi et al. 2001; Fukui et al. 2008).  
Higher resolution (FWHM $\sim$ 45'') observations of many of the regions detected in CO, using the Swedish-ESO Submillimetre (SEST) and Mopra Telescopes, have revealed additional structure on smaller spatial scales (Israel et al. 1993, 2003; Muller et al. 2010; Wong et al. 2011).
In the LMC and especially the SMC, the CO emission is typically both weaker than for comparably sized Galactic molecular clouds and more concentrated into colder, denser clumps -- with little CO emission from any more diffuse 'interclump' gas (e.g., Rubio et al. 1993a, 1993b; Lequeux et al. 1994).
Comparisons with maps of the 21 cm emission from \mbox{H\,{\sc i}} and the IR emission from dust, which is thought to trace both atomic and molecular gas, however, strongly suggest that there is a significant amount of more diffuse molecular gas present outside the molecular cores seen in CO (e.g., Leroy et al. 2007, 2009; Bernard et al. 2008; Bolatto et al. 2011).
Observations of additional molecular species, obtained for several regions where the CO emission is relatively strong, have yielded relative molecular abundances a factor of 10 or more smaller than those found (for example) in Orion KL and TMC-1 (Johansson et al. 1994; Chin et al. 1998).

Far-UV spectra covering the Lyman and Werner bands of H$_2$ were obtained (under various programs) for nearly 300 LMC and SMC sight lines with the {\it Far-Ultraviolet Spectroscopic Explorer} ({\it FUSE}) satellite, and are now available via the {\it FUSE} Magellanic Clouds Legacy Project (Blair et al. 2009).
Analysis of the H$_2$ absorption in $\sim$145 (mostly lightly reddened) sight lines has indicated both generally lower molecular fractions [$f$(H$_2$) = 2$N$(H$_2$)/$N$(H$_{\rm tot}$), where $N$(H$_{\rm tot}$) = $N$(\mbox{H\,{\sc i}})+2$N$(H$_2$)] and generally enhanced H$_2$ rotational excitation in the Magellanic Clouds -- requiring both a lower H$_2$ formation rate (presumably due to the lower dust-to-gas ratios) and increased photodissociation and excitation by the stronger ambient radiation fields (Tumlinson et al. 2002; Cartledge et al. 2005; Welty et al. 2012).
The kinetic temperatures inferred from the relative populations in the lowest two H$_2$ rotational levels -- generally between 40 and 120 K -- are similar to those found for diffuse molecular clouds in the local Galactic ISM, however.
And despite the differences in average total gas-to-dust ratios $N$(H$_{\rm tot}$)/$E(B-V)$ between the Milky Way, LMC, and SMC, the corresponding $N$(H$_2$)/$E(B-V)$ ratios are more similar in the three galaxies (Welty et al. 2012).

Observations of optical/UV absorption from other simple molecular species in the Magellanic Clouds have been much more limited.
CH and/or CH$^+$ have been detected toward three SMC and nine LMC stars; CN has been detected toward Sk~143 (SMC) and Sk$-$67~2 (LMC); and CO absorption has been reported toward three LMC stars (Magain \& Gillet 1987; Andr\'{e} et al. 2004; Welty et al. 2006; Cox et al. 2007).
The CH/H$_2$ ratio is comparable to that found for diffuse Galactic molecular clouds in some of the LMC and SMC sight lines, but is lower by factors up to 10--15 in others (Welty et al. 2006).
The CO column densities in the three LMC sight lines are slightly {\it higher} than for Galactic sight lines with comparable $N$(H$_2$) (Andr\'{e} et al. 2004).
The abundances of CH and CO thus appear to depend sensitively on local physical conditions -- not just on metallicity.
For the sight lines where CN is not detected, much of the CH (and perhaps CO) in diffuse molecular gas in the LMC and SMC may be produced via the still undetermined process(es) responsible for the observed CH$^+$ -- perhaps in more extended photon-dominated regions (Zsarg\'{o} \& Federman 2003; Welty et al. 2006).

In this paper, we present the first detections of absorption from C$_2$ and C$_3$ in the ISM of the SMC -- which are also the first such detections in any extragalactic system.
Section~\ref{sec-obsred} describes the observations and the processing of the spectroscopic data.
Section~\ref{sec-disc} discusses the derived abundances and rotational excitation of the two molecules.
Analysis of the rotational excitation of C$_2$ yields estimates for the temperature and density -- independent of those obtained from H$_2$ or detailed chemical models -- in the main SMC cloud containing the molecules.
Section~\ref{sec-summ} summarizes our results.

\begin{figure}
\includegraphics[width=84mm]{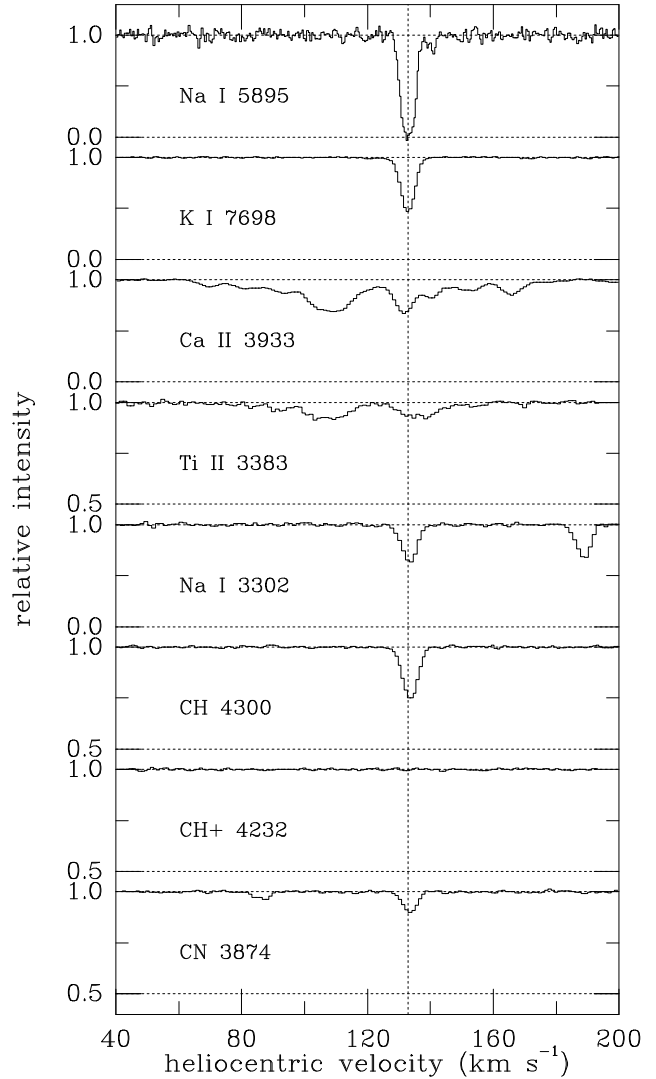}
\caption{Spectra of interstellar atomic and molecular absorption lines toward Sk 143 (SMC components only).
The \mbox{Na\,{\sc i}} $\lambda$5895 spectrum was obtained with the ESO CES (FWHM $\sim$ 1.35 km~s$^{-1}$; Welty \& Crowther, in prep.); all other spectra were obtained with the ESO UVES (FWHM $\sim$ 4 km~s$^{-1}$).
While many components are seen in \mbox{Ca\,{\sc ii}} and \mbox{Ti\,{\sc ii}}, the neutral atomic and molecular species are found predominantly in a single narrow component at about 132 km~s$^{-1}$.}
\label{fig:spec}
\end{figure}

\section{OBSERVATIONS AND DATA ANALYSIS}
\label{sec-obsred}

The sight line toward the moderately reddened [$E(B-V)_{\rm SMC}$ = 0.33] O9.7~Ib star Sk~143 (AzV 456), located in the near 'wing' region of the SMC, appears to be rather atypical of the SMC (within the current relatively small SMC samples).
The UV extinction curve for Sk~143 has a noticeable 2175 \AA\ bump and moderate far-UV rise -- more similar to typical Galactic extinction curves than those found for most other sight lines in the SMC (Lequeux et al. 1982; Gordon et al. 2003; though see Ma\'{i}z Apell\'{a}niz \& Rubio 2012). 
The gas-to-dust ratio, as given by $N$(H$_{\rm tot}$)/$E(B-V)$ = 8.2 $\times$ 10$^{21}$ cm$^{-2}$ mag$^{-1}$, is much closer to the average Galactic value (5.6 $\times$ 10$^{21}$ cm$^{-2}$ mag$^{-1}$) than to the average SMC value (23.4 $\times$ 10$^{21}$ cm$^{-2}$ mag$^{-1}$) (Welty et al. 2012).
The fraction of hydrogen in molecular form, $f$(H$_2$) $\sim$ 0.63, is the highest known in the SMC (Cartledge et al. 2005; Welty et al. 2012), and the depletion of titanium is the most severe (Welty \& Crowther 2010).
The relative abundances of \mbox{Na\,{\sc i}}, \mbox{K\,{\sc i}}, CH, CN, and the (unidentified) carriers of certain of the diffuse interstellar bands are more similar to those seen in the local Galactic ISM than in other SMC sight lines (Welty et al. 2006; Cox et al. 2007; Welty \& Crowther, in prep.); the \mbox{K\,{\sc i}}/\mbox{Na\,{\sc i}} ratio is unusually high.
The spectra discussed in this paper have also yielded the first detections of absorption from \mbox{Li\,{\sc i}} beyond our Galaxy (Howk et al. 2012).
While multiple SMC components are seen for \mbox{Ca\,{\sc ii}} and \mbox{Ti\,{\sc ii}}, the neutral atomic and molecular species are found predominantly in a single narrow ($b$ $\sim$ 0.5--0.8 km~s$^{-1}$) component near a heliocentric velocity of 132 km~s$^{-1}$ (Fig.~\ref{fig:spec}) -- which corresponds to the strongest component seen in \mbox{H\,{\sc i}} 21 cm emission (Welty et al. 2006, 2012; Cox et al. 2007; Howk et al. 2012).
The UV extinction, gas-to-dust ratio, and abundances seen toward Sk~143 do not seem to be characteristic of the entire SMC wing region, however (Welty et al. 2006, 2012; Cox et al. 2007; Welty \& Crowther 2010 and in prep.).

Optical spectra of Sk~143 were obtained with the ESO VLT Ultraviolet and Visual Echelle Spectrograph (UVES; Dekker et al. 2000) via service observing in 2008 September.
A standard dichroic \#2 setting, with grating central wavelengths near 3900 \AA\ (blue) and 7600 \AA\ (red), yielded spectral coverage from about 3290--4520 \AA\ in the blue and from about 5685--7525 and 7660--9465 \AA\ in the red.
While the primary aim of the observations was to obtain sensitive measurements of the interstellar \mbox{Li\,{\sc i}} lines near 6707 \AA\ (Howk et al. 2012), this setting also covers numerous other interstellar atomic and molecular absorption lines -- including the A-X (2-0) and (3-0) bands of C$_2$ and the $\tilde{A}$-$\tilde{X}$ (000-000) band of C$_3$ discussed in this paper.
Ten 2850-second exposures were taken through a 0.7 arcsec slit, and the resulting spectra were reduced via the UVES pipeline. 
The final summed spectra have a resolution of about 4.0 km~s$^{-1}$ (near 6700 \AA; Howk et al. 2012) and S/N ratios ranging from about 80 (per half resolution element) near 3300 \AA\ to about 300 near 4200 \AA.
Normalized spectral segments containing the various interstellar absorption features were obtained by dividing the summed spectra by low-order polynomial fits to the adjacent continuum regions (e.g., Fig.~\ref{fig:spec}).

\begin{table}
\caption{Sk 143:  SMC Equivalent Widths} 
\label{tab:skew}
\begin{tabular}{@{}lllr}
\hline
\multicolumn{1}{c}{Species}&
\multicolumn{1}{c}{Transition}&
\multicolumn{1}{c}{$\lambda$(\AA)}&
\multicolumn{1}{c}{W$_{\lambda}$(m\AA)}\\
\hline
CH      & B-X (0,0) $R_2$(1)                          & 3878.768 &   2.5$\pm$0.2 \\ 
        & B-X (0,0) $Q_2$(1)+$^{Q}R_{12}$(1)          & 3886.410 &   7.2$\pm$0.3 \\ 
        & B-X (0,0) $^{P}Q_{12}$(1)                   & 3890.213 &   4.8$\pm$0.4 \\ 
        & A-X (0,0) $R_{2e}$(1)+$R_{2f}$(1)           & 4300.3132&  21.7$\pm$0.4 \\ 
CH$^+$  & A-X (0,0) $R$(0)                            & 4232.548 &$<$0.7         \\ 
CN      & B-X (0,0) $R_1$(1)+$R_2$(1)+$^{R}Q_{12}$(1) & 3873.999 &   2.6$\pm$0.3 \\ 
        & B-X (0,0) $R_1$(0)+$^{R}Q_{12}$(0)          & 3874.607 &   7.0$\pm$0.3 \\ 
        & B-X (0,0) $P_1$(1)+$^{P}Q_{12}$(1)          & 3875.764 &   1.0$\pm$0.3 \\ 
C$_2$   & A-X (2,0) $R$(4)                            & 8751.685 &   2.4$\pm$0.9 \\
        & A-X (2,0) $R$(2)                            & 8753.949 &  10.5$\pm$1.3 \\
        & A-X (2,0) $R$(0)                            & 8757.686 &   6.6$\pm$1.1 \\
        & A-X (2,0) $Q$(2)                            & 8761.194 &   7.4$\pm$1.2 \\
        & A-X (2,0) $Q$(4)                            & 8763.751 &   5.7$\pm$1.0 \\
\hline
\end{tabular}
\medskip
~ ~ \\
Uncertainties are 1$\sigma$; limits are 3$\sigma$.
\end{table}

Table~\ref{tab:skew} lists the equivalent widths measured from the normalized spectra, for unblended lines of various molecular species seen toward Sk~143.
The uncertainties include contributions from both photon noise and continuum placement (Jenkins et al. 1973; Sembach \& Savage 1992).
For CH, CH$^+$, and CN, the equivalent widths generally are consistent with those derived from earlier, lower S/N UVES spectra of Sk~143 (Welty et al. 2006; Cox et al. 2007) -- but with a reduced upper limit for $N$(CH$^+$).

\begin{figure*}
\begin{minipage}{180mm}
\includegraphics[width=45mm,angle=-90.0]{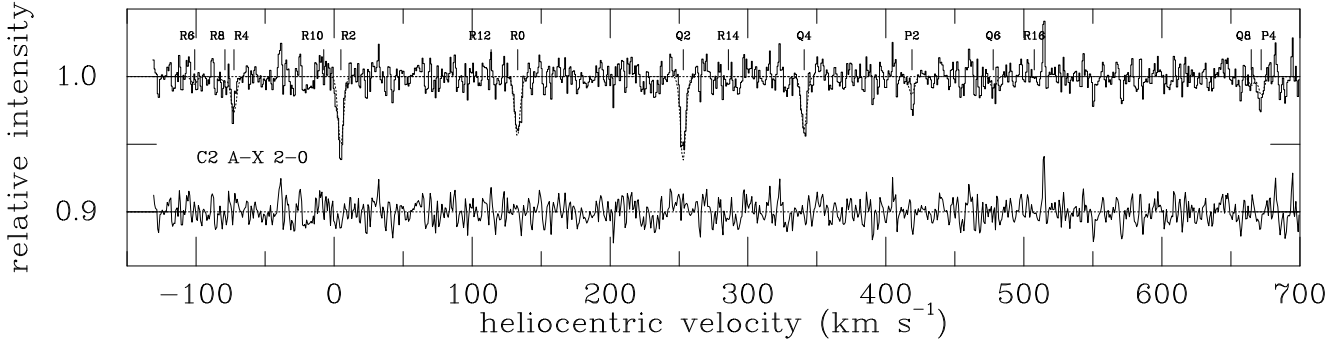}
\caption{C$_2$ A-X (2-0) band absorption (near 8760 \AA) toward Sk 143.
Lines from the P, Q, and R branches, from rotational levels $J$=0--14, are marked.
The velocity scale is referenced to the R0 line ($\lambda_{\rm rest}$ = 8757.686 \AA); the SMC absorption is at about 132 km~s$^{-1}$.
The smooth dotted line shows the simultaneous fit to the P, Q, and R branch lines from $J$=0--8.
The residual spectrum is shown below (centered at 0.9).}
\label{fig:c2sk}
\end{minipage}
\end{figure*}

\begin{figure*}
\begin{minipage}{180mm}
\includegraphics[width=45mm,angle=-90.0]{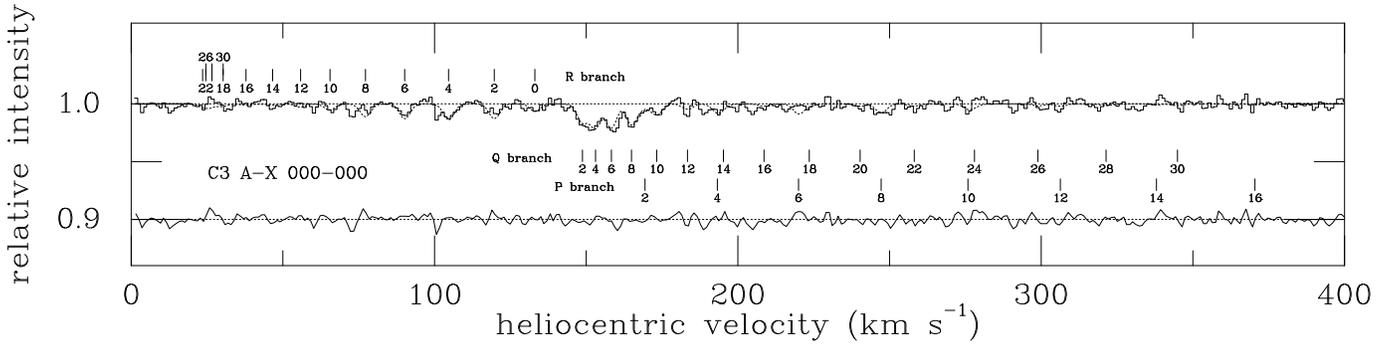}
\caption{C$_3$ $\tilde{A}$-$\tilde{X}$ (000)-(000) band absorption (near 4051 \AA) toward Sk 143.
Lines from the P, Q, and R branches, from rotational levels $J$=0--30, are marked.
The velocity scale is referenced to the R0 line ($\lambda_{\rm rest}$ = 4051.250 \AA); the SMC absorption is at about 132 km~s$^{-1}$.
The smooth dotted line shows the simultaneous fit to the P, Q, and R branch lines from $J$=0--12.
The residual spectrum is shown below (centered at 0.9).}
\label{fig:c3sk}
\end{minipage}
\end{figure*}

\begin{table}
\caption{Sk~143:  C$_2$ and C$_3$ Rotational Populations} 
\label{tab:rot}
\begin{tabular}{@{}rrrrr}
\hline
\multicolumn{1}{c}{$J$}&
\multicolumn{1}{c}{[$E$($J$)/$k$](C$_2$)}&
\multicolumn{1}{c}{$N_J$(C$_2$)}&
\multicolumn{1}{c}{[$E$($J$)/$k$](C$_3$)}&
\multicolumn{1}{c}{$N_J$(C$_3$)}\\
\multicolumn{1}{c}{ }&
\multicolumn{1}{c}{(K)}&
\multicolumn{1}{c}{(10$^{12}$ cm$^{-2}$)}&
\multicolumn{1}{c}{(K)}&
\multicolumn{1}{c}{(10$^{12}$ cm$^{-2}$)}\\
\hline
 0 &   0.0 &  7.2$\pm$1.2 &   0.0 & 0.16$\pm$0.07 \\
 2 &  15.6 & 20.9$\pm$2.1 &   3.7 & 0.84$\pm$0.16 \\
 4 &  52.1 & 13.0$\pm$2.2 &  12.4 & 1.10$\pm$0.12 \\
 6 & 109.4 &  2.9$\pm$1.2 &  26.1 & 1.11$\pm$0.16 \\
 8 & 187.6 &  2.0$\pm$1.2 &  44.7 & 1.00$\pm$0.14 \\
10 & 286.5 &    --        &  68.2 & 0.48$\pm$0.13 \\
12 & 406.3 &    --        &  96.7 & 0.25$\pm$0.12 \\
14 & 546.8 &    --        & 130.2 &  $<$0.26 \\
 & \\
$N$(obs) & & 46.0$\pm$3.6 &       & 4.94$\pm$0.35 \\
$N$(tot) & &   47$\pm$4   &       &  5.7$\pm$0.6 \\
\hline
\end{tabular}
\medskip
~ ~ \\
$N_J$ were determined using $f$-values from Sonnentrucker et al. (2007) for C$_2$ and from \'{A}d\'{a}mkovics et al. (2003) for C$_3$.
Uncertainties are 1$\sigma$, and reflect uncertainties in equivalent widths; limits are 3$\sigma$.
Total column densities include estimates for unobserved higher $J$ levels, based on derived or assumed excitation temperatures.
\end{table} 

\begin{table} 
\caption{Sk 143:  SMC Column Densities} 
\label{tab:skcd}
\begin{tabular}{@{}lcc}
\hline
\multicolumn{1}{c}{Species}&
\multicolumn{1}{c}{log[$N$ (cm$^{-2}$)]}&
\multicolumn{1}{c}{Ref}\\
\hline
\mbox{H\,{\sc i}} &   21.03$\pm$0.04 & 1,2 \\ 
H$_2$             &   20.93$\pm$0.09 & 3 \\ 
CH                &   13.54$\pm$0.04 & 4 \\ 
CH$^+$            &$<$11.88          & 4 \\
CN                &   12.42$\pm$0.03 & 4 \\
C$_2$             &   13.67$\pm$0.04 & 4 \\
C$_3$             &   12.76$\pm$0.05 & 4 \\
\hline
\end{tabular}
\medskip
~ ~ \\
References: 1 = Welty et al. 2012; 2 = Howk et al. 2012; 3 = Cartledge et al. 2005; 4 = this paper [using $f$-values from Gredel et al. (1991, 1993) for CH, CH$^+$, and CN, from Sonnentrucker et al. (2007) for C$_2$, and from \'{A}d\'{a}mkovics et al. (2003) for C$_3$]
\end{table}

Column densities for the molecular species were determined from the measured equivalent widths (for weak, unblended lines), by integration of the apparent optical depth over the absorption-line profiles, and via detailed fits to the line profiles (using the program {\sc fits6p}; e.g., Welty, Hobbs, \& Morton 2003).
Figures~\ref{fig:c2sk} and \ref{fig:c3sk} show the normalized spectra of the C$_2$ A-X (2-0) band near 8760 \AA\ (with S/N $\sim$ 140) and the C$_3$ $\tilde{A}$-$\tilde{X}$ (000)-(000) band near 4051 \AA\ (with S/N $\sim$ 270), respectively, toward Sk~143.
Absorption from rotational levels $J$ = 0--8 (for C$_2$) and $J$ = 0--12 (for C$_3$) can be discerned in the spectra at a velocity of about 132 km~s$^{-1}$; for C$_2$, it is clear that the bulk of the absorption is in levels $J$ = 0--4.
Column densities for the individual C$_2$ and C$_3$ rotational levels (Table~\ref{tab:rot}) were determined via simultaneous single-component fits to the lines from the P, Q, and R branches, using the rest wavelengths and $f$-values tabulated by Sonnentrucker et al. (2007) for C$_2$ and by \'{A}d\'{a}mkovics, Blake, \& McCall (2003) for C$_3$. 
A $b$-value of 0.7 km~s$^{-1}$, required to obtain consistent column densities from both weak and strong lines of \mbox{K\,{\sc i}} and \mbox{Na\,{\sc i}} (Welty et al. 2006; Howk et al. 2012; this study), was adopted for the lines of CN, C$_2$, and C$_3$ as well.
Because the absorption from C$_2$ and C$_3$ is relatively weak [e.g., $<$11 m\AA\ for the individual C$_2$ A-X (2-0) lines], the column densities are insensitive to the adopted $b$-value, for $b$ $>$ 0.3 km~s$^{-1}$.
Because the $b$-values were fixed in fitting the profiles, the uncertainties derived in those fits for the individual $N$($J$) were underestimated.
The somewhat larger values listed in Table~\ref{tab:rot} reflect the uncertainties in the equivalent widths of the corresponding absorption lines.
The adopted fits to the C$_2$ and C$_3$ bands are shown by the smooth dotted curves in Figs.~\ref{fig:c2sk} and \ref{fig:c3sk}, respectively.
The residuals (data minus fit) are shown below the spectra (centered at 0.9) in each figure.
The C$_2$ column densities derived from the A-X (2-0) band yield an acceptable fit to the weaker A-X (3-0) band.
The total $N$(C$_2$) = 4.7$\pm$0.4 $\times$ 10$^{13}$ cm$^{-2}$ and $N$(C$_3$) = 5.7$\pm$0.6 $\times$ 10$^{12}$ cm$^{-2}$ listed in the table include contributions from unobserved higher rotational levels, based on analyses of the rotational excitation (see below).
For C$_2$, the unobserved levels contribute less than 2 per cent of the total.
For C$_3$, however, the typical shallow fall-off of the higher-$J$ populations (e.g., \'{A}d\'{a}mkovics et al. 2003) suggests that the unobserved levels may contribute of order 15 per cent of the total column density [based on an assumed excitation temperature of 400 K for $J$ $\ge$ 14 (typical of the sight lines in \'{A}d\'{a}mkovics et al. 2003), with $N$($J$=14) set by an extrapolation of the lower-$J$ populations]. 
The molecular column densities, together with those of \mbox{H\,{\sc i}} and H$_2$ (Cartledge et al. 2005; Welty et al. 2012; Howk et al. 2012), are listed in Table~\ref{tab:skcd}.
The uncertainties listed for the total column densities include the uncertainties in the contributions from those unobserved levels.

\section{RESULTS / DISCUSSION}
\label{sec-disc}

\subsection{Abundances of C$_2$ and C$_3$}
\label{sec-abund}

The chemistry and excitation behaviour of C$_2$ in diffuse and translucent clouds have been explored in a number of previous studies (e.g., van Dishoeck \& Black 1982; van Dishoeck \& de Zeeuw 1984; Federman \& Huntress 1989; Federman et al. 1994; Sonnentrucker et al. 2007).
In such clouds, the route to C$_2$ formation begins with the reaction C$^+$~+~CH~$\rightarrow$~C$_2^+$~+~H, followed by a series of hydrogen abstraction reactions and dissociative recombinations which yield C$_2$ via several channels.
At low-to-moderate optical depths and densities, the destruction of C$_2$ is dominated by photodissociation.
For Galactic clouds with total visual extinction $A_{\rm V}$ $\la$ 5 mag, the models of van Dishoeck \& Black (1989) predict a roughly constant C$_2$/H$_2$ ratio $\sim$ 3.3--5.7 $\times$ 10$^{-8}$.
Because C$_2$ is a 'second generation' molecule -- dependent on the prior existence of H$_2$ and CH -- it is likely to be found primarily in somewhat colder, denser regions.

\begin{figure*}
\begin{minipage}{180mm}
\includegraphics[width=160mm]{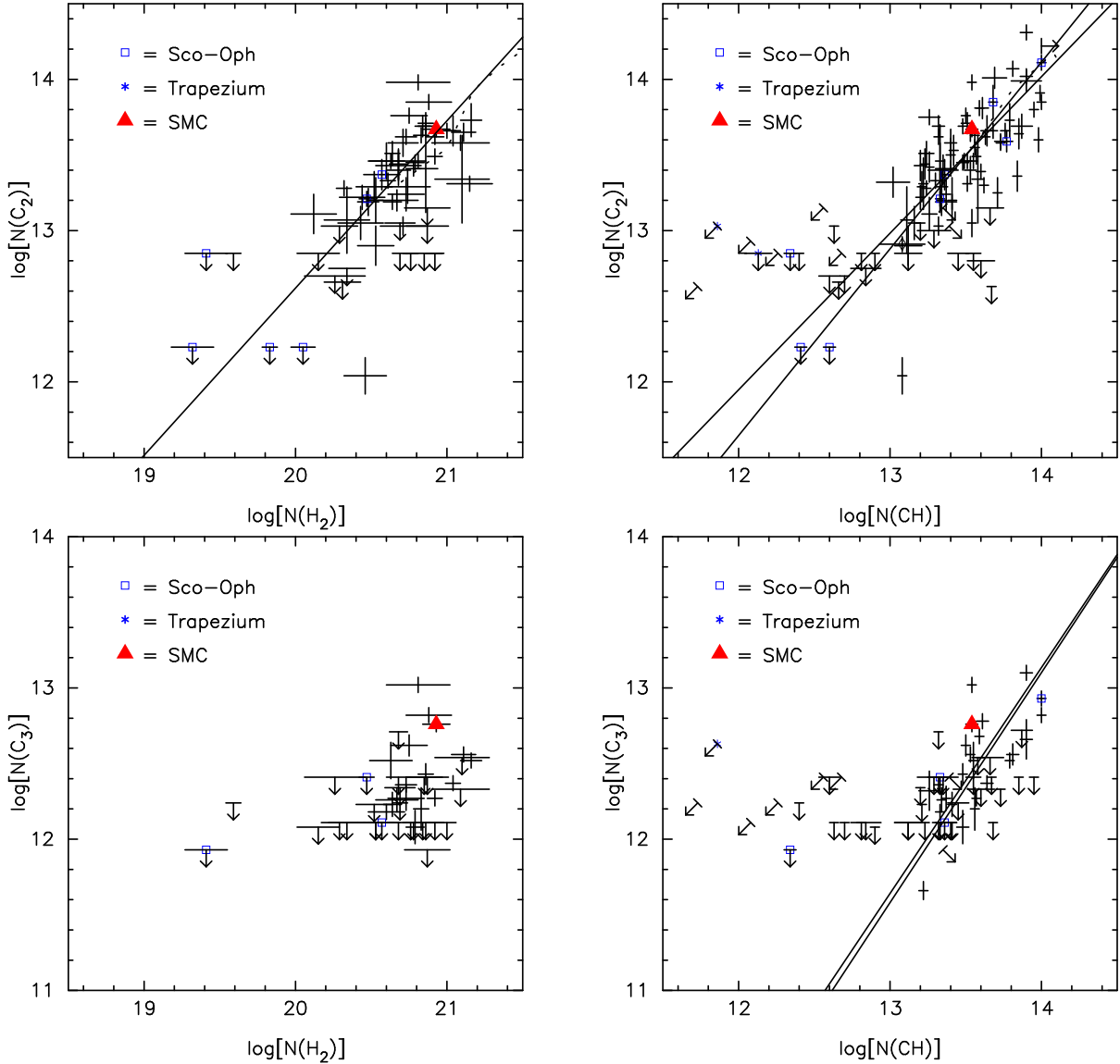}
\caption{$N$(C$_2$) ({\it upper}) and $N$(C$_3$) ({\it lower}) vs. $N$(H$_2$) ({\it left}) and $N$(CH) ({\it right}).
Red triangle denotes Sk 143; blue open squares denote Sco-Oph sight lines; plain crosses denote other Galactic sight lines.
Straight solid lines show weighted and unweighted fits to the Galactic data; dotted curves for C$_2$ vs. H$_2$ and CH show predictions from models T1--T6 of van Dishoeck \& Black (1988).} 
\label{fig:corr1}
\end{minipage}
\end{figure*}

The upper left panel of Figure~\ref{fig:corr1} shows the relationship between the column densities of C$_2$ and H$_2$, for Galactic sight lines (black crosses) and for the SMC gas toward Sk~143 (red triangle).
As $N$(CH) is both generally well correlated with $N$(H$_2$) and known for some sight lines with no data for H$_2$, the upper right panel shows the corresponding relationship between $N$(C$_2$) and $N$(CH).
The Galactic C$_2$ data are from a new, expanded survey of C$_2$ absorption (Welty, Black, \& McCall, in prep.; see also Sonnentrucker et al. 2007).
The observed $N$(C$_2$) and $N$(H$_2$) are moderately correlated, with linear correlation coefficient $r$ = 0.57, but there are some sight lines with weaker than 'expected' C$_2$.
While the slightly steeper than linear relationship between the two column densities now agrees reasonably well with the predicted trend (dotted curve), the sample exhibits rather limited ranges in the column densities of the two molecules.
A slightly steeper than linear relationship, with $r$ = 0.66, is also seen for the somewhat larger sample of $N$(C$_2$) versus $N$(CH).
The C$_2$/H$_2$ ratio toward Sk~143, 5.5 $\times$ 10$^{-8}$, is quite consistent with the Galactic values (as was found for CH and CN in that sight line; Welty et al. 2006) -- despite the significantly lower overall metallicity in the SMC.

The sample of C$_3$ detections in the Galactic ISM is still relatively small (Maier et al. 2001; Roueff et al. 2002; Galazutdinov et al. 2002; Oka et al. 2003; \'{A}d\'{a}mkovics et al. 2003), and its abundance and rotational excitation are not as well understood as those of C$_2$.
Examination of gas-phase chemical networks suggests that most C$_3$ is formed via dissociative recombination of C$_3$H$^+$ -- though neutral-neutral reactions (e.g., C + C$_2$H$_2$) may also contribute (Roueff et al. 2002).
The immediate precursor C$_3$H$^+$ may be produced via several pathways involving various ion-neutral reactions (e.g., beginning with ionization of C$_2$ and involving intermediary species such as C$_3^+$, C$_2$H$_2$, C$_2$H, and C$_2$H$_3^+$; Oka et al. 2003).
The destruction of C$_3$ is also likely dominated by photodissociation.

The two lower panels of Figure~\ref{fig:corr1} show the relationships between $N$(C$_3$) and the column densities of H$_2$ and CH.
The relatively small sample size and limited range in the observed $N$(C$_3$) make it difficult to characterize those relationships, and there are some sight lines with upper limits for C$_3$ that are well below the fitted trend with $N$(CH).
It does appear, however, that the C$_3$ abundance toward Sk~143 is entirely consistent with the Galactic values as well.

Given the significantly lower average metallicity in the SMC (and the even lower carbon abundance; e.g., Garnett 1999; Trundle \& Lennon 2005), we might have expected the abundances of various carbon-containing molecules to be correspondingly lower there than in Galactic diffuse molecular clouds -- as is found for CH toward the SMC stars Sk~13, Sk~18, and AzV~476 (Welty et al. 2006).
The higher abundances of CH, CN, C$_2$, and C$_3$ observed toward Sk~143, which are similar to local Galactic values (Welty et al. 2006; Cox et al. 2007; this paper), are thus rather puzzling.

One possible explanation for the higher molecular abundances -- a higher than typical metallicity in that sight line -- may be suggested by the gas-to-dust ratio toward Sk~143, which is 3--4 times lower than the average SMC value (and only slightly higher than the local Galactic average; Welty et al. 2012).
Interstellar metallicities may be estimated by comparing the column densities of dominant, little-depleted species (e.g., \mbox{O\,{\sc i}}, \mbox{S\,{\sc ii}}, \mbox{Zn\,{\sc ii}}) with the total hydrogen column density.
For example, toward Sk~78, Sk~108, and Sk~155 -- three SMC sight lines with fairly well determined $N$(\mbox{Zn\,{\sc ii}}) -- the observed log[$N$(\mbox{Zn\,{\sc ii}})/$N$(H$_{\rm tot}$)] range from $-$8.04 to $-$8.24 dex (Welty et al. 2001, 2012, and in prep.).
Those interstellar zinc abundances are consistent with an overall SMC zinc abundance of $-$8.0 dex [a factor of 4.3 lower than the solar abundance; from analyses of SMC Cepheids by Luck \& Lambert (1992) and Luck et al. (1998)] and zinc depletions of less than a factor of 2 (similar to the depletions seen in the Galactic ISM; e.g., Jenkins 2009).
Unfortunately, the available STIS spectra of the \mbox{Zn\,{\sc ii}} lines at 2026 and 2062 \AA\ toward Sk~143 are rather noisy (Sofia et al. 2006; Howk et al. 2012), and the strongest \mbox{Zn\,{\sc ii}} component (near 132 km~s$^{-1}$) is somewhat saturated.
Apparent optical depth integrations over the \mbox{Zn\,{\sc ii}} line profiles (Howk et al. 2012) indicate that the zinc depletion is less than a factor of 4 (for the adopted SMC zinc abundance), but the gas-phase \mbox{Zn\,{\sc ii}} abundance could be much higher.

While strong constraints thus cannot yet be placed on the interstellar metallicity toward Sk~143, we believe that it is unlikely that the metallicity could be as high as solar there. 
In an early survey of abundances in 27 SMC \mbox{H\,{\sc ii}} regions, Pagel et al. (1978) found an essentially constant oxygen abundance O/H = 7.98$\pm$0.09 dex (a factor of about 5 below solar), with very little scatter and no discernible radial gradient or regional trends.
More recent studies of abundances in SMC field B stars (e.g., Dufton et al. 2005; Trundle \& Lennon 2005) indicate that O, Mg, and Si are sub-solar by 0.6--0.8 dex and that C is sub-solar by 1.1 dex (all with rms scatter less than 0.2 dex); similar results are obtained for B stars in the clusters NGC~330 and NGC~346 (Hunter et al. 2009).
While there are individual stars with somewhat higher reported carbon abundances (for example), those values generally are based on measurements of a single line, and the abundances of the other elements in those stars are not similarly enhanced.
The existing nebular and stellar abundance studies thus offer little evidence for significantly higher than average metallicities in individual regions or sight lines in the SMC.

Another way of increasing the gas-phase carbon abundance toward Sk~143 -- less severe depletion of carbon into dust grains -- also seems unable to account for the higher molecular abundances there.
While there are indications of milder than 'expected' depletions of Mg, Si, and Ti in several SMC sight lines, and while the overall depletions of titanium are generally much less severe in the SMC than in Galactic clouds (Welty et al. 2001 and in prep.; Welty \& Crowther 2010; cf. Sofia et al. 2006), the sight line toward Sk~143 exhibits the most severe titanium depletion in the current (fairly small) SMC sample -- so it seems unlikely that carbon would be significantly less severely depleted there.
Moreover, carbon is typically depleted by less than a factor of 3 in the Galactic ISM (e.g., Jenkins 2009), so that even a complete lack of depletion could not increase the carbon abundance by the necessary factor of 4--5 (or 8--10, in view of the lower overall SMC carbon abundance).

Alternatively, there might be some structural or environmental factors which affect the chemistry in such a way as to enhance the carbon-containing diatomics (and C$_3$) toward Sk~143.
For example, Welty et al. (2006) conjectured that the Galactic-like CH abundances in several LMC sight lines might be due to nonthermal formation of CH, together with CH$^+$, in more extensive photon-dominated regions (e.g., Zsarg\'{o} \& Federman 2003).
Since CH is a chemical predecessor of C$_2$, enhanced CH might thus lead to a correspondingly higher C$_2$ abundance.
That scenario is not likely to apply toward Sk~143, however.
CN is detected there, and CH$^+$ is not (Table~\ref{tab:skcd}) -- so that the CH is more likely to be a product of equilibrium chemistry. 
Moreover, the temperature and density derived from C$_2$ rotational excitation (see next section) suggest that the C$_2$ is found in a cooler, denser part of the cloud than the H$_2$ (and, presumably, the CH).

The relatively high density inferred from C$_2$ (next section) suggests another explanation for the high molecular abundances.
For densities less than several thousand per cm$^3$, models of steady-state gas phase chemistry predict that the CH abundance will be determined by a balance between formation (initiated by the reaction C$^+$ + H$_2$ $\rightarrow$ CH$_2^+$ + $h\nu$) and photodissociation.
The CH/H$_2$ ratio thus will be proportional to both the carbon abundance and the ratio $n_{\rm H}$/$I_{\rm UV}$ (e.g., eqn.~4 in Welty et al. 2006), where $I_{\rm UV}$ gives the strength of the ambient UV radiation field (in units of the average Galactic field).
The ratio $n_{\rm H}$/$I_{\rm UV}$ may be obtained from the column densities of \mbox{H\,{\sc i}} and H$_2$, using estimates for the rate coefficients for H$_2$ formation ($R$) and photoabsorption ($\beta_0$) (e.g., Lee et al. 2002; Welty et al. 2006):
\begin{equation}
\frac{n_{\rm H}}{I_{\rm UV}} = \frac{(0.11)\beta_0(8.45\times10^5)}{R} \frac{[N({\rm H}_2)]^{1/2}}{N({\rm H}_{\rm tot})}.
\end{equation}
In the Galactic ISM, most sight lines with N(H$_2$) $\ga$ 10$^{19}$ cm$^{-3}$ have $n_{\rm H}$/$I_{\rm UV}$ in the range 5--25 cm$^{-3}$ (using $R$ = 3 $\times$ 10$^{-17}$ cm$^3$s$^{-1}$).
For sight lines in the Magellanic Clouds, $R$ will be somewhat smaller, due (at least in part) to the lower dust-to-gas ratios (e.g., Welty et al. 2012); Tumlinson et al. (2002) adopted $R$ = 3 $\times$ 10$^{-18}$ cm$^3$s$^{-1}$.
With that choice for $R$, most of the Magellanic Clouds sight lines with $N$(H$_2$) $\ga$ 10$^{19}$ cm$^{-2}$ (Tumlinson et al. 2002; Cartledge et al. 2005; Welty et al. 2012) have somewhat higher $n_{\rm H}$/$I_{\rm UV}$, in the range 20--80 cm$^{-3}$.
The sight line to Sk~143 has the highest $n_{\rm H}$/$I_{\rm UV}$ ratio in the current SMC sample, at about 165 cm$^{-3}$ -- roughly an order of magnitude higher than the typical Galactic values.
[Given the higher dust-to-gas ratio toward Sk~143 (Welty et al. 2012), however, $R$ may also be higher there -- and the inferred $n_{\rm H}$/$I_{\rm UV}$ ratio correspondingly lower.
On the other hand, $I_{\rm UV}$ (which is independent of $R$) does seem to be somewhat low (for the SMC) toward Sk~143 (Welty et al. 2006), and $n_{\rm H}$ somewhat high (see next section) -- so that the $n_{\rm H}$/$I_{\rm UV}$ ratio may still be enhanced in that sight line.]
A high $n_{\rm H}$/$I_{\rm UV}$ ratio toward Sk~143 thus may offset the low SMC carbon abundance, yielding a CH abundance comparable to that found in the Galactic ISM.
For the two other SMC sight lines with CH detections (Sk~18, AV~476), both the $n_{\rm H}$/$I_{\rm UV}$ ratios and the CH/H$_2$ ratios are significantly lower (Welty et al. 2006).
Where CH is enhanced, other molecules formed via reactions involving CH (e.g., C$_2$) may also exhibit higher abundances.

\subsection{Rotational excitation of C$_2$ and C$_3$}
\label{sec-tex}

Analysis of the rotational excitation of C$_2$ can yield estimates for both the kinetic temperature and the density in diffuse molecular clouds (e.g., van Dishoeck \& Black 1982; Sonnentrucker et al. 2007).
In such clouds, most of the C$_2$ is in various rotational levels of the ground electronic and vibrational state.
Levels in higher electronic states are populated primarily by absorption of near-IR photons [to $A~^1\Pi_u$ (Phillips system)].
The molecule then cascades back to rotational levels in the ground vibrational state via quadrupole and intersystem (singlet-triplet) transitions.
Collisions with H and H$_2$ then can modify those lower-level populations (van Dishoeck \& Black 1982; van Dishoeck \& de Zeeuw 1984).
For moderate densities, the resulting populations of the excited rotational levels are higher than their thermal equilibrium values -- with corresponding excitation temperatures greater than the actual kinetic temperature ($T_{\rm k}$).
In practice, the relative populations in the lowest rotational levels provide the best constraints on $T_{\rm k}$, while the populations in the higher rotational levels yield constraints on the local density of collision partners [$n_{\rm c}$ = $n$(H) + $n$(H$_2$)].
For Galactic sight lines with well-determined C$_2$ rotational populations (e.g., up to $J$ = 12), the inferred $T_{\rm k}$ is typically 20--50 K, while $n_{\rm c}$ is typically 150--350 cm$^{-3}$ (Sonnentrucker et al. 2007; Welty et al., in prep.).

\begin{figure}
\includegraphics[width=84mm]{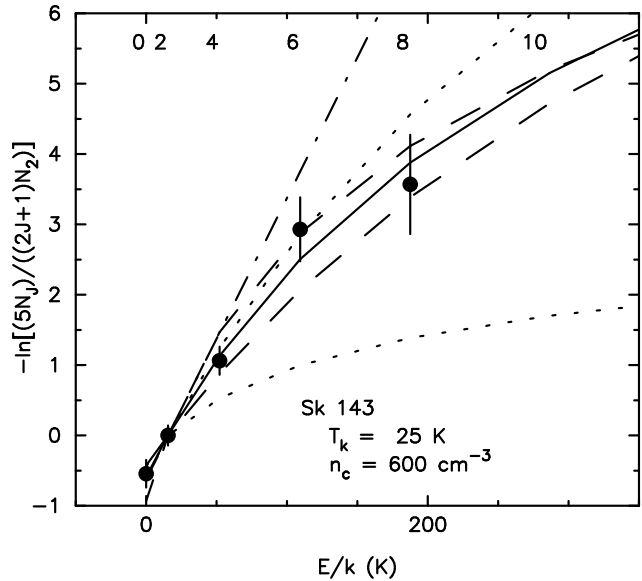}
\caption{Rotational excitation of C$_2$ toward Sk~143, relative to populations in $J$ = 2, as a function of excitation energy.
The solid line shows the best-fitting theoretical curve, for $T_k$ = 25 K and $n_c$ = 600 cm$^{-3}$ (assuming a typical {\it Galactic} near-IR radiation field).
The straight dot-dashed line shows the relative populations for thermal equilibrium at that $T_k$; the two dotted curves show the relative populations for that $T_k$ and $n_c$ = 1000 (nearer the equilibrium line) and 100 cm$^{-3}$.
The two dashed lines show the relative populations for ($T_k - 10$,$n_c - 50$) and ($T_k + 10$, $n_c + 50$).}
\label{fig:c2rot}
\end{figure}

Figure~\ref{fig:c2rot} shows the normalized excitation diagram ($-$ln[(5~$N_{J}$)/((2$J$+1)~$N_2$)] vs. $E_{J}$/k) for C$_2$ toward Sk~143.
The observed relative C$_2$ rotational populations are compared with those predicted by the models of van Dishoeck \& Black (1982; see also van Dishoeck \& de Zeeuw 1984; Sonnentrucker et al. 2007) -- in order to estimate both the actual kinetic temperature and the density.
The models may be parametrized by $n_{\rm c}\sigma$/$I$, where $n_{\rm c}$ is the density of collision partners; $\sigma$ is the collisional de-excitation cross-section for transitions from $J+2$ to $J$ (assumed 2$\times$10$^{-16}$ cm$^{2}$); and $I$ is a scaling factor for the strength of the interstellar radiation field in the near-IR (i.e., near the A-X bands).
Higher values of $n_{\rm c}\sigma/I$ yield predicted rotational populations closer to the thermal equilibrium values (dot-dashed line in the figure).
The solid curve shows the populations predicted for the best-fitting $T_{\rm k}$ $\sim$ 25 K and $n_{\rm c}$ $\sim$ 600 cm$^{-3}$ (assuming a typical Galactic near-IR field).
Slightly poorer, but still acceptable fits to the rotational population distribution are obtained for 20 K $\la$ $T_{\rm k}$ $\la$ 30 K and $n_{\rm c}$ $\ga$ 400 cm$^{-3}$.
That range in $T_{\rm k}$ is quite consistent with the excitation temperatures derived from the lowest two, three, or four rotational levels:  $T_{02}$ = 29$\pm$6 K, $T_{04}$ = 33$\pm$1 K, $T_{06}$ = 32$\pm$1 K.
For $f$(H$_2$) = 0.63, the best-fitting total hydrogen density $n$(\mbox{H\,{\sc i}}) + 2$n$(H$_2$) is about 870 cm$^{-3}$ (with a lower limit of about 580 cm$^{-3}$) -- much higher than the 85 cm$^{-3}$ estimated from the excitation of H$_2$ (Welty et al. 2006)\footnotemark, and also higher than the values typically found in the local Galactic ISM (Sonnentrucker et al. 2007; Welty et al., in prep.).
\footnotetext{From H$_2$ rotational excitation, Welty et al. (2006) found the UV radiation field toward Sk~143 to be slightly weaker than the Galactic average -- but that was based on a crudely estimated column density for H$_2$ $J$=4.
Re-examination of the {\it FUSE} spectra of Sk~143 suggests that $N$($J$=4) could be somewhat higher -- so that $I_{\rm UV}$ and $n_{\rm H}$ could be as high as $\sim$ 2 and $\sim$ 300 cm$^{-3}$, respectively.
Because the C$_2$ rotational excitation depends on $n_{\rm c}\sigma$/$I$, the derived densities should be scaled by the actual field strength (in the near-IR).}
For such high densities, there is very little C$_2$ in the higher rotational levels -- less than 2 per cent of the total C$_2$ for the level populations calculated for the best-fitting $T_{\rm k}$ $\sim$ 25 K and $n_{\rm c}$ $\sim$ 600 cm$^{-3}$.

\begin{figure}
\includegraphics[width=84mm]{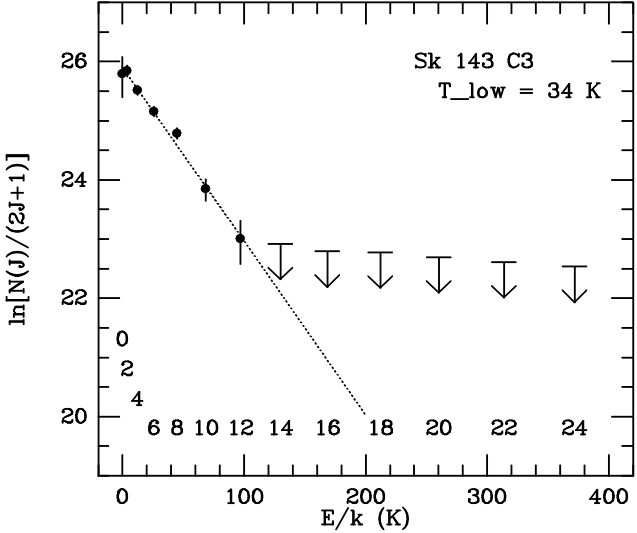}
\caption{Rotational excitation of C$_3$ toward Sk~143, as a function of excitation energy.
The straight dotted line is a linear fit to the (normalized) populations for $J$ = 0--12.
The slope of that line corresponds to a temperature $T_{\rm low}$ of about 34 K.}
\label{fig:c3rot}
\end{figure}

Similar considerations apply to the rotational excitation of C$_3$, where higher ro-vibrational levels (bending and streching modes) may be populated via IR and far-IR radiation, with subsequent cascades back to rotational levels in the ground electronic state.
At densities below about 10$^3$ cm$^{-3}$, both the IR radiation field and collisions with H$_2$ (and H?) will affect the rotational excitation.
Both observations of the C$_3$ $\tilde{A}$-$\tilde{X}$ (000)-(000) band (\'{A}d\'{a}mkovics et al. 2003) and theoretical models of C$_3$ excitation (Roueff et al. 2002) suggest that the lower C$_3$ rotational levels ($J$ = 0--12 or so) will be characterized by a fairly low excitation temperature (slightly higher than the kinetic temperature), while the higher C$_3$ rotational levels will exhibit somewhat higher excitation temperatures.

Figure~\ref{fig:c3rot} shows the normalized C$_3$ rotational populations ln[$N$($J$)/(2$J$+1)] observed toward Sk~143.
The populations in $J$ = 0--12 are well-fitted with a single excitation temperature $T_{\rm ex}$ $\sim$ 34 K, which lies between the kinetic temperatures inferred from C$_2$ (25 K) and H$_2$ (45 K).
Unfortunately, the lack of data for higher rotational levels precludes a full analysis of the excitation.
In order to estimate the total C$_3$ column density, we have adopted an excitation temperature of 400 K for $J$ $\ge$ 14 (typical of the sight lines analyzed by \'{A}d\'{a}mkovics et al. 2003), with $N$($J$=14) set by an extrapolation of the lower-$J$ populations.
The lack of a visible R-branch band head (which would be near 25 km~s$^{-1}$ in Fig.~\ref{fig:c3sk}) suggests that the higher-$J$ populations could not contribute much more than the resulting 15 per cent of the total $N$(C$_3$).

Both the differences in kinetic temperature and density inferred from C$_2$ and H$_2$ and the rather high density inferred from C$_2$ may reflect several factors.
$T_{\rm k}$(C$_2$) is often smaller than T$_{01}$(H$_2$) in Galactic sight lines, suggesting that C$_2$ may generally be concentrated in colder, denser regions than H$_2$ (e.g., Sonnentrucker et al. 2007).
Moreover, theoretical models of interstellar clouds predict that higher pressures/densities are needed for a stable cold, neutral phase in lower-metallicity systems like the SMC (due to reduced cooling, which is normally dominated by \mbox{C\,{\sc ii}} fine-structure emission; e.g., Wolfire et al. 1995).
Thermal pressures estimated for several SMC sight lines from the fine-structure excitation of \mbox{C\,{\sc i}} appear to be consistent with such predictions (Welty et al., in prep.).
Diffuse molecular clouds in the lower-metallicity SMC may have (relatively) more extensive outer zones, where H$_2$ is self-shielded but other molecules are still strongly subject to photodissociation (e.g., Pak et al. 1998), and the density contrast between the inner and outer zones might be more extreme.
While a higher $N$[H$_2$($J$=4)] (see footnote 1) would increase the density inferred from H$_2$, it would also imply a stronger UV radiation field.
If the near-IR field were correspondingly stronger, then the density inferred from C$_2$ would also be higher.

The high density implied by the excitation of C$_2$ may also be partly an artefact of the outdated molecular data embedded in the analysis originally proposed by van Dishoeck and Black (1982). 
Although the more recent analyses of C$_2$ column density and hydrogen number density have been scaled for improved values of oscillator strengths in the A-X, D-X, and F-X electronic transitions, they have not incorporated recent improvements in collisional excitation rates. 
The downward rates for inelastic collisions of C$_2$ with He (Najar et al. 2008) and ground-state H$_2$ (Najar et al. 2009) are larger than and vary with $J$ differently from the rates based on the simple scaling law adopted by van Dishoeck and Black (1982). 
A partial update of the old model suggests that the inferred density toward Sk~143 has been slightly overestimated, but not by a factor as large as two; similar analyses of Galactic sight lines would also be affected.
Casu \& Cecchi-Pestellini (2012) have computed models of C$_2$ excitation based on new molecular data. 
There is no simple trend in inferred density; rather, Casu \& Cecchi-Pestellini (2012) contend that existing observations of diffuse molecular clouds are best fitted by inhomogeneous models with both dense and diffuse phases. 
Much more work on the excitation models is needed: reliable collision rates for ortho-H$_2$ and atomic H do not exist and chemical effects have not yet been taken into account. 

\section{SUMMARY} 
\label{sec-summ}

We have discussed the first detection of absorption from interstellar C$_2$ and C$_3$ beyond our Galaxy -- toward Sk 143, located in the near 'wing' region of the SMC.
The total abundances of C$_2$ and C$_3$ (relative to H$_2$) are similar to those found in diffuse Galactic molecular clouds -- consistent with previous results for CH, CN, and several diffuse interstellar bands -- despite the significantly lower overall metallicity of the SMC.
While a higher metallicity toward Sk~143 cannot be completely ruled out, we believe that the enhanced molecular abundances are much more likely to reflect environmental factors -- in particular the higher than usual $n_{\rm H}$/$I_{\rm UV}$ ratio -- affecting the chemistry in that SMC sight line.
Analysis of the rotational excitation of C$_2$ ($J$ = 0--8) yields an estimated kinetic temperature $T_{\rm k}$ $\sim$ 25 K and a total hydrogen density $n_{\rm H}$ $\sim$ 870 cm$^{-3}$ -- compared to the $T_{01}$ $\sim$ 45 K and $n_{\rm H}$ $\sim$ 85--300 cm$^{-3}$ obtained from H$_2$.
The populations of the lower rotational levels of C$_3$ ($J$ = 0--12) are consistent with an excitation temperature of about 34 K.
The differences in $T_{\rm k}$ and density inferred from C$_2$ and H$_2$ may reflect differences in the distribution of the two molecular species within the main cloud/component in this sight line; such differences may be more pronounced at low metallicities.
The relatively high density inferred from C$_2$ (compared to values found for Galactic clouds) is qualitatively consistent with theoretical predictions for cold, neutral clouds in lower metallicity systems. 




\end{document}